\documentclass[preprint,prd,showpacs,floatfix]{revtex4}
\usepackage{amsmath}
\usepackage{amsfonts}
\usepackage{amssymb}
\usepackage{graphicx}

\begin{document}

\title{Non-Cancellation of Quantum Geometry Fluctuations}

\author{Victor Parkinson }
\email{victor@cosmos.phy.tufts.edu}
\author{L. H. Ford }
\email{ford@cosmos.phy.tufts.edu}
\affiliation{Institute of Cosmology, Department of Physics and Astronomy\\
    Tufts University, Medford, MA 02155, USA}
    
\begin{abstract}
Quantum vacuum fluctuations tend to be strongly anti-correlated, which reduces their
observable effects. However, time dependence can upset the cancellation of these
anti-correlated fluctuations and greatly enhance their effects. This form of non-cancellation
is investigated for spacetime geometry fluctuations driven by the vacuum stress tensor fluctuations of the
electromagnetic field. The time dependence can take the form of sinusoidal conformal
metric perturbations or of time variation of the gravitational constant, both of which can arise
from modifications of the gravitational action. We examine two observable quantities, luminosity fluctuations
and redshift fluctuations, both of which can arise from  stress tensor fluctuations. We find
that both quantities can grow with increasing distance between a source and an observer. This
secular growth raises the possibility of observing  spacetime geometry fluctuation effects at length 
scales far longer than the Planck scale.
\end{abstract}

\maketitle

\section{Introduction}
\label{sec:intro}

Classical general relativity is a geometric theory of gravity in which the effects of gravity
upon light rays play a crucial role. Thus gravitational lensing, gravitational redshift, and
Shapiro time delay~\cite{Shapiro} are well-tested predictions of the classical theory. 
In any quantum description
of gravity, quantum fluctuations in light propagation are expected, and can be regarded as 
lightcone fluctuations. Pauli~\cite{Pauli} was one of the first to recognize the significance of lightcone
fluctuations in quantum gravity. Numerous authors in recent years have discussed various
aspects of this phenomenon~\cite{F95,FS96,Moffat,HS98,EMN99,BF04,BF04b,TF06,YSF09,Ng}. 
The search for potentially observable effects from fluctuations of
gravity is the phenomenology of quantum gravity, and is reviewed in Ref.~\cite{A-C-review}.
Papers on this topic can be roughly divided into two categories: those which seek predictions
from  particular proposal for full quantum gravity, such as string theory or loop quantum gravity,
and those which restrict attention to weak quantum fluctuations, either from quantized linear
perturbations (active fluctuations) or from quantum stress tensor fluctuations (passive fluctuations).  
The present paper deals with the last topic, specifically the effects of vacuum fluctuations of the
electromagnetic field stress tensor on the propagation of light via passive fluctuations of a weak 
gravitational field. 
Quantum fluctuations of the stress tensor operator have been studied by many authors, and have
recently been reviewed in Refs.~\cite{FW07,Stochastic}.

Vacuum fluctuations can be exceedingly hard to observe because of their strong
anti-correlations. This can be illustrated by vacuum fluctuations of the electric field. If these
fluctuations were uncorrelated, then a charged particle would undergo Brownian motion and
acquire energy which is not available in the vacuum state. Although the particle can temporarily
receive energy from a vacuum fluctuation, an anti-correlated fluctuation will take the energy
back on a time scale consistent with the energy-time uncertainty principle~\cite{YF04}.  
However, if there is a source of time dependence, and hence an external source of energy, then 
the anti-correlations can be upset and the particle can gain energy~\cite{BF09,PF11}. For example,
Ref.~\cite{PF11} deals with a charged particle oscillating near a mirror. It is possible to arrange the
parameters so that the mean squared velocity of the particle grows linearly in time, for a finite period.
The oscillatory motion of the charge upsets the cancellation of the anti-correlated fluctuations,
and allows the particle to undergo Brownian motion. This charged particle model is a useful
analog model for the quantum gravitational effects which will be discussed in the present paper.
In particular,  quantum stress tensors exhibit similar anti-correlation behavior~\cite{FR05}.

We will investigate two effects which can arise from Riemann tensor fluctuations driven by quantum
stress tensor fluctuations. The first effect is fluctuations in the observed luminosity of a source viewed
through a fluctuating spacetime geometry~\cite{BF04}. This effect is analogous to scintillation,
or ``twinkling" of a source viewed though a medium with fluctuating density. The second effect is
broadening of a spectral line from a source~\cite{TF06}. This effect can be viewed as due to fluctuations
in the gravitational redshift in a fluctuating gravitational field. Both effects can be calculated from
integrals of the  Riemann tensor correlation function. We consider only passive fluctuations driven
by a quantum stress tensor, so this correlation function can be expressed in terms of the elctromagnetic stress tensor
correlation function. In both cases, we use a geometric optics approximation in which the fluctuating
geometry is probed by light rays whose local wavelength is short compared to the scale associated
with the time variation of the background. 
In Sec.~\ref{sec:lum}, the luminosity fluctuations are calculated for a spatially flat universe with sinusoidal
scale factor oscillations and are found to grow with increasing distance. The origin of such oscillations
from modifications of Einstein's equations is discussed in Sec.~\ref{sec:osc}. The redshift fluctuations 
in the same universe are calculated in Sec.~\ref{sec:red}, and a similar growth with distance is found.
The possibility of obtaining both effects from an oscillating gravitational constant is considered in
Sec.~\ref{sec:var-G}. The results of the paper are summarized and discussed in Sec.~\ref{sec:sum}.

Units in which $\hbar =c =1$ are employed throughout this paper, so Newton's constant becomes
$G= \ell_p^2$, where $\ell_p$ is the Planck length.

\section{Luminosity Fluctuations}
\label{sec:lum}

\subsection{Basic Formalism}
\label{sec:basic}

In this section, we discuss fluctuations in the apparent luminosity of a distant source,
due to Ricci tensor fluctuations driven by quantum stress tensor fluctuations. The basic
formalism which we need was developed in Ref.~\cite{BF04}, and will be briefly reviewed
and adapted to the present problem. The key idea is to treat the Raychaudhuri equation
as a Langevin equation. Consider a bundle of null geodesics with expansion $\theta$,
affine parameter $\lambda$, and tangent vector $k^\mu = dx^\mu/d\lambda$. If the vorticity
vanishes, and the shear and squared expansion may be neglected, then the Raychaudhuri 
equation becomes
\begin{equation}
\frac{d\theta}{d\lambda} = -R_{\mu \nu}(x)k^{\mu}(x)k^{\nu}(x)\,,
\label{eq:Ray}
\end{equation}
where $R_{\mu \nu}$ is the Ricci tensor. We may integrate this equation along a portion of
a geodesic, and then calculate the variance of the expansion as a double integral:
\begin{equation}
\langle (\Delta \theta)^2\rangle =  \int  d \lambda_1 \, d \lambda_2 \: K_{\mu \nu \alpha \beta}(x_1,x_2)\,
k^{\mu}(x_1) \,k^{\nu}(x_1) \, k^{\alpha}(x_2) \, k^{\beta}(x_2)  \,.
\label{eq:Ricci-int}
\end{equation}
Here 
\begin{equation}
K_{\mu \nu \alpha \beta}(x_1,x_2) =  
\langle R_{\mu \nu}(x_1)R_{\alpha \beta}(x_2)\rangle - 
\langle R_{\mu \nu}(x_1)\rangle \langle R_{\alpha \beta}(x_2)\rangle 
\label{eq:Ricci-corr}
\end{equation}
is Ricci tensor correlation function. If the stress tensor is traceless, the Ricci tensor is given by
\begin{equation}
R_{\mu\nu} = 8\pi\ell_p^2 T_{\mu\nu}\,,
\end{equation}
and hence
\begin{equation}
K_{\mu \nu \alpha \beta}(x_1,x_2) =  64 \pi^2 \ell_p^4\, C_{\mu \nu \alpha \beta}(x_1,x_2)\,
\end{equation}
where $C_{\mu \nu \alpha \beta}(x_1,x_2)$ is the stress tensor correlation function. Note that the 
renormalized expectation value, $ \langle T_{\mu \nu} \rangle$, may have a nonzero trace due to the
conformal anomaly, but this trace cancels in the correlation function, and can be neglected here.

We are interested in the case where the stress tensor fluctuations induce passive spacetime geometry
fluctuations around a mean Robertson-Walker spacetime, which we take to be spatially flat and
described by a metric of the form
\begin{equation}
ds^2 =  -dt^2 +a^2( dx^2  +dy^2 +dz^2) = a^2(\eta)(-d\eta^2 + dx^2  +dy^2 +dz^2) \,,
\label{eq:metric}
\end{equation}
where $t$ is comoving time, $\eta$ is conformal time, and $a$ is the scale factor. We take our null
rays to be moving in the $x$-direction, and define null coordinates by
\begin{equation}
\begin{split}
u_1 = & \: \eta_1 - x_1 \\
u_2 = & \: \eta_2 - x_2 \\
v_1 = & \: \eta_1 + x_1 \\
v_2 = & \: \eta_2 + x_2 \,,
\end{split}
\end{equation}
where the subscripts refer to the two spacetime points in the double integration in Eq.~(\ref{eq:Ricci-corr}).
The source and the detector are taken to be at rest, and have the four-velocity 
\begin{equation}
t^\alpha = a^{-1}\, \delta^\alpha_\eta 
\label{eq:t}
\end{equation}
The null vector is found from the geodesic equation to be
\begin{equation}
k^{\beta}(\eta_1) = a^{-2}(\eta_1) \, \delta^\beta_{v_1} \,.
\label{eq:k}
\end{equation}
Consequently, the affine parameter is related to the advance time null coordinate by 
$d\lambda_1 = a^{2}(\eta_1) \, dv_1$

\begin{figure}
  \centering
  \scalebox{.5}{\includegraphics[scale=1]{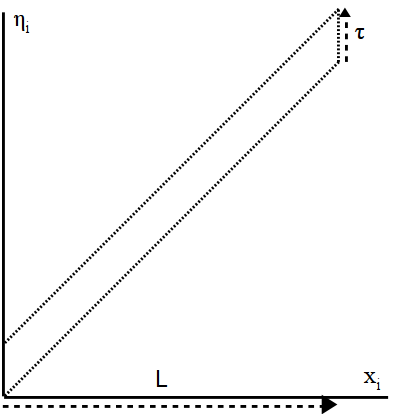}}\\
  \caption{The parallelogram shows the spacetime region defined by the bundle of light rays.
  The source and detector are separated by a coordinate distance $L$, and the pulse has a
  temporal duration of $\tau$ in the rest frame of both the source and the detector.}
  \label{fig:integ_area1}
\end{figure}

\begin{figure}
  \centering
  \scalebox{.5}{\includegraphics[scale=1]{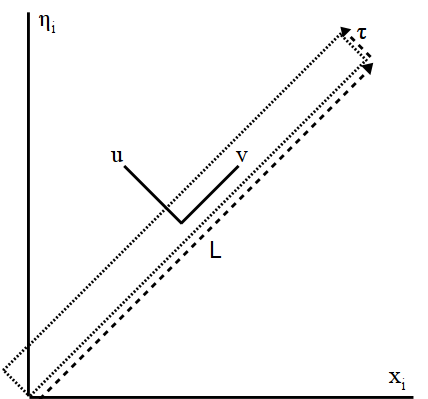}}\\
  \caption{The spacetime region of integration has been approximated by the rectangle
  illustrated here. The variables of integration has been changed to null coordinates $u$ and $v$. }
  \label{fig:integ_area2}
\end{figure}

The quantum stress tensor correlation function is singular in the coincidence limit, but its integrals
can be finite, so it is a meaningful distribution. However, in our problem it is necessary to integrate
over  a temporal direction, as well as along the null geodesic. We can take this to be an average
over the time of emission or detection of the photons. In this case, we will be integrating over the
parallelogram region depicted in Fig.~\ref{fig:integ_area1}. 
It will be convenient to change the integration variables from
$(\eta_1,v_1,\eta_2,v_2)$ to  $(u_1,v_1,u_2,v_2)$. Further, the length of the interval, $L$, in $v_i$
is much longer than the length, $\tau$, in $v_i$, so we may approximate the integration region with the
rectangle shown in Fig.~\ref{fig:integ_area2}.
 Now our expression for the variance of the expansion becomes
\begin{equation}
\begin{split}
\langle (\Delta \theta)^2\rangle = 
{} & 64 \pi^2 \ell^4_p \int du_1\, f(u_1,\tau) \int du_2\, f(u_2,\tau) \int_0^{L} dv_1 \int_0^{L} dv_2 \\
& \times a^3(\eta_1) a^3(\eta_2)\, C_{\mu \nu \alpha \beta}k^{\mu} k^{\nu}k^{\alpha} k^{\beta} \,
\end{split} \label{eq:-C-int}
\end{equation}
where $f(u_i)$ is a sampling function of width $\tau$. We obtain two powers of the scale factor in each
variable from the change of variables from $\lambda_i$ to $v_i$ and one power from changing the temporal
integration variable from $t_i$ to $\eta_i$. We also use the relation $d\eta_i = du_i$ when $x_i$
is constant.

We will adopt a Lorentzian form for the sampling function:
\begin{equation}
 f(u_i,\tau) = \frac{\tau}{\pi(u_i^2 + \tau^2)} \, , 
\end{equation}
so that 
\begin{equation}
\int_{-\infty}^{\infty} du_i f(u_i,\tau) = 1\,.
\end{equation}
It will also be convenient to replace the step functions in $v_1$ and $v_2$ of width $L$, which appear
in Eq.~(\ref{eq:-C-int}) by Lorentzian  functions with the same area. That is, we will replace
\begin{equation}
\int_{0}^{L} dv_i \rightarrow L \int_{-\infty}^{\infty} dv_i \; f(v_i,L)  = L\,.
\label{eq:replace}
\end{equation}
Now Eq.~(\ref{eq:-C-int}) can be written as
\begin{equation}
\begin{split}
\langle (\Delta \theta)^2\rangle = 
{} & 64 \pi^2 \ell^4_p \, L^2 \int_{-\infty}^{\infty} du_1\, f(u_1,\tau) \int _{-\infty}^{\infty}du_2\, f(u_2,\tau) 
\int_{-\infty}^{\infty} dv_1 f(v_1,L)  \int_{-\infty}^{\infty} dv_2 f(v_2,L)  \\
& \times a^3(\eta_1) a^3(\eta_2)\, C_{\mu \nu \alpha \beta}k^{\mu} k^{\nu}k^{\alpha} k^{\beta} \,
\end{split} \label{eq:-C-int2}
\end{equation}

The stress tensor correlation function in the Robertson-Walker spacetime of Eq.~(\ref{eq:metric}) can be
obtained by a conformal transformation from flat spacetime. Recall that the comoving energy density scales
as $a^{-4}$, so $T_{tt} = a^{-2}\, T_{\eta\eta} = a^{-4}\, T^{flat}_{tt}$ so  we have
\begin{equation}
C_{\mu \nu \alpha \beta}(x_1,x_2) =  a^{-2}(\eta_1)\, a^{-2}(\eta_2)\,  
C^{flat}_{\mu \nu \alpha \beta}(x_1,x_2)\,.
\label{eq:weight}
\end{equation}
The flat space correlation function for the electromagnetic field was given in Ref.~ \cite{FW03} as
\begin{equation} \label{eq:EM}
\begin{split}
C^{flat}_{\mu \nu \alpha \beta}(x_1,x_2) = {} & 4(\partial_{\mu} \partial_{\nu}D)(\partial_{\alpha}\partial_{\beta}D) \\
& + 2g_{\mu\nu}(\partial_{\alpha}\partial_{\rho}D)(\partial_{\beta}\partial^{\rho}D) + 2g_{\alpha\beta}(\partial_{\mu}
\partial_{\rho}D)(\partial_{\nu}\partial^{\rho}D) \\
& - 2g_{\mu\alpha}(\partial_{\nu}\partial_{\rho}D)(\partial_{\beta}\partial^{\rho}D) - 2g_{\nu\alpha}(\partial_{\mu}
\partial_{\rho}D)(\partial_{\beta}\partial^{\rho}D) \\
& - 2g_{\nu\beta}(\partial_{\mu}\partial_{\rho}D)(\partial_{\alpha}\partial^{\rho}D)- 2g_{\mu\beta}(\partial_{\nu}
\partial_{\rho}D)(\partial_{\alpha}\partial^{\rho}D) \\
& + (g_{\mu\alpha}g_{\nu\beta} + g_{\nu\alpha}g_{\mu\beta} - g_{\mu\nu}g_{\alpha\beta})(\partial_{\rho}\partial_{\sigma}D)
(\partial^{\rho}\partial^{\sigma}D) \,.
\end{split}
\end{equation}
Here $D$ is the two-point function for a massless scalar field $\varphi$ given by
\begin{equation}
D(x_1,x_2) = \langle \varphi(x_1) \varphi(x_2) \rangle =  \frac{1}{4\pi^2(x_1 - x_2)^2}
=  \frac{1}{4\pi^2\,[(\mathbf{x}_1 - \mathbf{x}_2)^2 - (\eta_1 - \eta_2 - i\epsilon)^2] } \,.
\end{equation}
Here the $ - i\epsilon$ serves as a convergence factor in the mode sum, which contains a term of the
form $\exp[-i \omega ( \eta_1 - \eta_2)]$, and will be valuable in defining integrals of the correlation function.

The contracted correlation function which appears in Eq.~(\ref{eq:-C-int}) may be written as
\begin{equation} 
 C_{\mu \nu \alpha \beta}k^{\mu} k^{\nu}k^{\alpha} k^{\beta} = 
 \frac{4}{a^6(\eta_1)a^6(\eta_2)}(\partial^2_{v_1} D)(\partial^2_{v_2}D)\,,
\end{equation}
where we have used Eqs.~(\ref{eq:k}), (\ref{eq:weight}), and (\ref{eq:EM}). If we neglect the size of the bundle
of rays in the transverse spatial dimensions compared to the temporal duration, then we can write the two-point 
function as 
\begin{equation}
D(x_1,x_2) = - \frac{1}{4\pi^2\,[ (u_1 - u_2 - i\epsilon)   (v_1 - v_2 - i\epsilon)] } \,,
\label{eq:D}
\end{equation}
and find
\begin{equation} 
 C_{\mu \nu \alpha \beta}k^{\mu} k^{\nu}k^{\alpha} k^{\beta} = 
\frac{1}{\pi^4\,a^6(\eta_1) \,a^6(\eta_2) \,(u_1-u_2- i\epsilon)^2\,(v_1-v_2- i\epsilon)^6}\,.
\end{equation}
This leads to 
\begin{equation}
\begin{split}
\langle (\Delta \theta)^2\rangle = 
{} & \frac{64 \ell^4_p}{\pi^2}\,L^2\, \int_{-\infty}^{\infty} du_1\, f(u_1,\tau) \int _{-\infty}^{\infty}du_2\, f(u_2,\tau) 
\int_{-\infty}^{\infty} dv_1f(v_1,L)  \int_{-\infty}^{\infty} dv_2 f(v_2,L)  \\
& \times a^{-3}(\eta_1) a^{-3}(\eta_2)\, \frac{1}{(u_1-u_2- i\epsilon)^2\,(v_1-v_2- i\epsilon)^6}  \,.
\end{split} \label{eq:-C-int3}
\end{equation}

\subsection{Scale Factor Oscillations}
\label{sec:osc}

Our model assumes that the scale factor is sinusoidally oscillating around flat spacetime with angular
frequency $\omega$ and amplitude $A \ll 1$, so we have
\begin{equation}
a(\eta) = 1 + A\sin(\omega\eta) \,.
\label{eq:a-osc}
\end{equation}
These oscillations are crucial in preventing cancellation of the anti-correlated fluctuations, and are
analogous to the oscillation of the charge or of the mirror assumed in Ref.~\cite{PF11}. In this subsection
we discuss how such oscillations  can arise in modifications of general relativity. Suppose that the Einstein 
action is altered by adding to  the scalar curvature $R$ a term proportional to $R^2$:
\begin{equation}
R \rightarrow R + \frac{1}{2}\, a_2\, R^2 \,.
\end{equation}
The Einstein equation is modified by the addition of a term to the stress tensor of the form
\begin{equation}
\frac{a_2}{16 \pi} \, I_{\mu\nu} = \frac{a_2}{16 \pi} \, \left( 2 \nabla_\mu \nabla_\nu R 
-2 g_{\mu \nu}  \nabla^\sigma \nabla_\sigma R - 2 R\, R_{mu\nu} +\frac{1}{2} g_{\mu \nu} \, R^2 \right)\,.
\end{equation}
The new term is fourth-order in derivatives of the metric and arises in the process of renormalizing the
expectation value of the quantum matter stress tensors. If $a_2 < 0$ then flat spacetime becomes unstable.
If $a_2 > 0$, then the effect of this term is to produce oscillations of the scale factor of the form of
Eq.~\eqref{eq:a-osc}, at a frequency of~\cite{HW78}
\begin{equation}
\omega = \frac{1}{\sqrt{3 \, a_2}} \,.
\label{eq:osc-freq}
\end{equation}
Note that this frequency depends inversely upon the square root of the constant $a_2$. However, the
observational constraints on this constant are not especially strong. The best upper bound comes from
laboratory tests of the inverse square law of gravity, and is about~\cite{BG11}
\begin{equation}
a_2 < 2 \times 10^{-9} {\rm m}^2 \,,
\end{equation}
leading to a lower bound on $\omega$ of
\begin{equation}
\omega > 1.3  \times 10^{4} {\rm m}^{-1} =  4 \times 10^{12} {\rm Hz}\,.
\end{equation}

Horowitz and Wald~\cite{HW78} have suggested that scale factor oscillations could cause radiation by
charged particles, leading to stronger constraints on $a_2$. This mechanism would presumably not
rule out values of $\omega$ greater than the particle rest mass, as charged particles cannot be treated
as classical point particles on scales approaching the Compton wavelength. This consideration would
also only apply to oscillations in the recent universe, and not rule out a period of oscillation in the early
universe. If energetic massless particles were to pass through this region of oscillation, fluctuation effects
from that era would in principle be observable much later.

\subsection{Calculations and Results}

Now we use Eq.~\eqref{eq:a-osc} for the scale factor in Eq.~(\ref{eq:-C-int3}). Because $A \ll 1$,
the part of the resulting expression which depends upon the scale factor may be Taylor expanded in powers
of $A$ as
\begin{equation}
\begin{split}
\frac{1}{a^3(\eta_1)a^3(\eta_2)} = {} &1 - 3 A \left[\sin \left(\eta _1 \omega \right)+\sin \left(\eta _2 \omega \right)\right]+ \\
&A^2 \left[6 \sin ^2\left(\eta _1 \omega \right)+ 6 \sin ^2\left(\eta _2 \omega \right)+9 \sin \left(\eta _2 \omega \right) \sin \left(\eta _1 \omega \right)\right] + \ldots \,.
\end{split}
\end{equation}
We terminate the expansion at second order, and rewrite the result in terms of complex exponentials as
\begin{equation}
\begin{split}
\frac{1}{a^3(\eta_1)a^3(\eta_2)} = {} & 1+ 6A^2   - \frac{3A}{2}ie^{-i{\omega}(u_1+v_1)/2} + \frac{3A}{2}ie^{i{\omega}(u_1+v_1)/2}- \frac{3A}{2}ie^{-i{\omega}(u_2+v_2)/2}\\
&   + \frac{3A}{2}ie^{i{\omega}(u_2+v_2)/2} - \frac{3A^2}{2} e^{-i\omega (u_1+v_1)}- \frac{3A^2}{2} e^{i\omega (u_1+v_1)}\\
& - \frac{3A^2}{2} e^{-i\omega (u_2+v_2)}- \frac{3A^2}{2} e^{i\omega (u_2+v_2)} 
  - \frac{9A^2}{4} e^{i \omega (u_1+u_2+v_1+v_2)/2}  \\
& - \frac{9A^2}{4} e^{-i {\omega}(u_1+u_2+v_1+v_2)/2}  + \frac{9A^2}{4} e^{-i{\omega}(u_1-u_2+v_1-v_2)/2}\\
& + \frac{9A^2}{4} e^{i {\omega}(u_1-u_2+v_1-v_2)/2} \,.
\end{split}
\label{eq:expand}
\end{equation}

The only term in the above expansion which yields a growing contribution is the last term. 
The constant term  gives a contribution to $\langle (\Delta \theta)^2\rangle$ proportional 
to the flat spacetime contribution, which does not grow. The remaining terms, other than the last, will also not grow,
as can be verified by explicit calculation. 
If we retain only the last term in Eq.~(\ref{eq:expand}), then the integrand in the expression for 
$\langle (\Delta \theta)^2\rangle$ is a function of $u=u_1-u_2$ and of $v=v_1-v_2$ only. This allows us to use
a property of Lorentzians:
\begin{equation}
\int_{-\infty}^{\infty} du_1 \int_{-\infty}^{\infty} du_2 f(u_1,\tau)f(u_2,\tau) F(u_1-u_2) =
 \int_{-\infty}^{\infty} du f(u,2\tau) F(u) \,,
\label{eq:lorentzian_id}
\end{equation}
and write
\begin{equation}
\langle (\Delta \theta)^2\rangle = \frac{36 \ell_p^4 A^2}{\pi^4}\int_{-\infty}^{\infty} du \int_{-\infty}^{\infty} dv 
\frac{\mu}{u^2 + \mu^2}\frac{\ell^3}{v^2 + \ell^2}\frac{e^{i \frac{\omega}{2}(u+v)}}{(u- i\epsilon)^2\,(v- i\epsilon)^6}\,,
\label{eq:lum_int}
\end{equation}
where $\mu=2\tau$ and $\ell=2L$.

The $v$-integration will be performed first. The integral of the $v$-dependent part of the integrand may be
defined to be
\begin{equation}
V = \int_{-\infty}^{\infty} dv \frac{\ell^3}{(v+i\ell)(v-i\ell)} \frac{e^{i \omega v/2}}{(v- i\epsilon)^6} \,.
\end{equation} 
The form of the exponential dictates that we should close the integration contour in the upper-half
$v$-plane, and enclose the simple pole at $v = i \ell$ and the sixth-order pole at $v = i\epsilon$. This leads
to the result, in the limit $\epsilon \rightarrow 0$,
\begin{equation}
V = - \frac{\pi }{1920 \ell^4}\; 
\left[ \ell \omega\; \left(\ell^4 \omega ^4+80 \ell^2 \omega ^2+1920\right)+1920e^{-\ell\omega/2} \right] 
\sim    -\frac{\pi  \omega^5}{1920} \; \ell \,,       
\end{equation}
where the last term is the asymptotic form for large $\ell$. This is the only term of interest, as it contributes
the part which grows with increasing distance. Note that it comes from the sixth-order pole. The simple
pole in the Lorenztian function has a contribution which decays exponentially with increasing distance. 
We will retain only the term proportional to $\ell$, and write
\begin{equation}
\langle (\Delta \theta)^2\rangle \sim -\ell\, \frac{3\ell_p^4 A^2 \omega^5}{160\pi^3}
 \int_{-\infty}^{\infty} du \frac{\mu}{u^2 + \mu^2} \frac{e^{i \omega u/2}}{(u- i\epsilon)^2} \,.
\end{equation}
As before, we close the contour in the upper-half $u$-plane, and enclose a simple pole at $u = i \ell$ 
and a second-order pole at $u = i\epsilon$. 

The result of this integration may be written in terms of $L$ and $\tau$ as
\begin{equation}
\langle (\Delta \theta)^2\rangle \sim 
\frac{3A^2 \omega^5 \ell_p^4 (e^{-\tau \omega}+2\tau \omega)}{320\pi^2\tau^2}\; L  \,.
 \end{equation}
This form holds when $L \gg 1/\omega$, and displays the linear growth in $L$. The associated variance
of the fractional luminosity is~\cite{BF04}
\begin{equation}
\left\langle \left( \frac{\Delta \mathcal{L}}{\mathcal{L}} \right)^2 \right\rangle = 
\frac{1}{4}L^2\langle (\Delta \theta)^2\rangle  \,.
\label{eq:lum}
\end{equation}
Strictly, this is the variation in the number luminosity, the number of photons per unit are arriving
at the source, and arises from the expansion fluctuations. Here we are ignoring any variations
in the energy per photon, which will be the topic of the next section.
Note that the root-mean-square  fractional luminosity grows as $L^{3/2}$. In the limit that 
$\tau \gg 1/\omega$, it may be expressed as
\begin{equation}
\left( \frac{\Delta \mathcal{L}}{\mathcal{L}} \right)_{rms} \sim
 \frac{1}{16\pi} \sqrt{\frac{6 L^3}{5 \tau}}  \ell_p^2 A \omega^3 \,.
\end{equation}
This  can be written as
\begin{equation}
\left( \frac{\Delta \mathcal{L}}{\mathcal{L}} \right)_{rms} \sim
6 \times 10^{-3} \, A\, \left(\frac{L}{\rm 1 Gpc}\right)^\frac{3}{2}  \, 
\left(\frac{\omega}{\rm 100 keV}\right)^3\, \left(\frac{\rm 1 s}{\tau}\right)^\frac{1}{2}\,.
 \label{eq:lum-est}
\end{equation}
The key feature of this result is that the growth with increasing distance $L$ has
created an effect which can be significant far from the Planck scale. 

It is useful to add some comments as to why only the last term in Eq.~(\ref{eq:expand}) yields
a contribution which grows with increasing $L$. As noted before, the constant terms have the same
form as in Minkowski spacetime, where anti-correlated fluctuations prevent growth. The next-to-last
term is evaluated by closing the contour in the lower-half $v$-plane avoiding the pole at $v = i\epsilon$,
and leaving only the residue of the pole at $v = -i \ell$, which decays exponentially.
The remaining terms in  Eq.~(\ref{eq:expand})  depend upon $v_1$ and $v_2$ in some combination 
other than their difference, $v$, which is the combination appearing in the denominator of 
Eq.~\eqref{eq:-C-int3}. An alternative way to evaluate the  $v_1$ and $v_2$ integrations is to change
variables to $v$ and $q = v_1 + v_2$. In this formulation, the term linear in $L$ can only arise from
the $q$ integral of an integrand which is independent of $q$, which does not arise for these remaining
terms.

\section{Redshift Fluctuations}
\label{sec:red}

\begin{figure}
  \centering
  \scalebox{.5}{\includegraphics[scale=1]{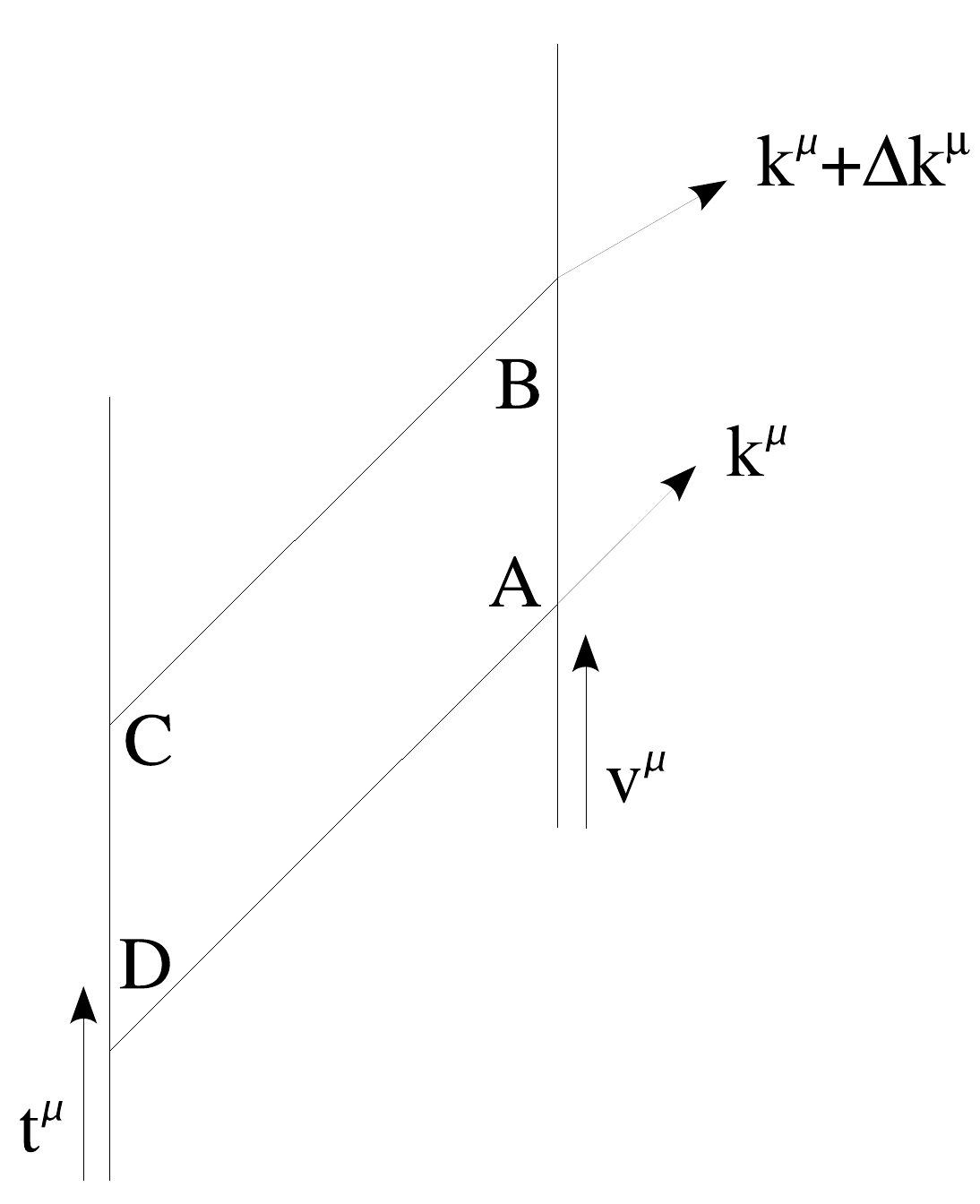}}\\
  \caption{A source moves along a worldline with tangent $t^{\mu}$
  while a detector a proper distance $s$ away moves along a worldline
  with tangent $v^{\mu}$. The source emits a ray  which has
  tangent $k^{\mu}(\lambda=0)$ at point D and tangent $k^{\mu}(\lambda_0)$ 
  at A.
  Parallel propagation of $k^{\mu}$
  around ABCD results in a slightly rotated vector $k^{\mu} + \Delta k^{\mu}$.
  The closed path ABCD encloses the spacetime region of interest.}
  \label{fig:spacetime}
\end{figure}

In addition to the luminosity fluctuations treated in the previous section, spacetime
geometry fluctuations can also induce broadening of spectral line through redshift
fluctuations. The formalism for describing this effect in terms of Riemann tensor
fluctuations was developed in Ref.~\cite{TF06}, and will be reviewed here. The basic
physical idea is that spacetime geometry fluctuations lead to variations in the gravitational
redshift between a source and a detector. Let $\xi = \Delta \nu/\nu$ be the
fractional redshift, where $\nu$ is the frequency in the frame of the source, and $\nu +\Delta \nu$
is that in the frame of the detector. The rate of change of $\xi$ in the frame of the detector may
be written as an integral of the Riemann tensor along the null geodesic connecting the
events of emission and detection as~\cite{BM71,TF06}
 \begin{equation}
 \frac{d\xi}{d\tau} = v_{\mu} \int_0^{\lambda_0}
 d\lambda \,
 R^{\mu}_{\phantom{\mu}\alpha\nu\beta}k^{\alpha}t^{\nu}k^{\beta}\,.
 \label{RateOfChange}
\end{equation}
Here $k^\mu$ is the tangent to the null geodesic, $t^\mu$ is the four-velocity of the source, and
$v^\mu$ is that of the detector, as illustrated in Fig.~\ref{fig:spacetime}.

Here we are concerned with changes in gravitational redshift, not with Doppler shifts due
to the relative motion of the source and detector. We are also interested in fluctuations
around a mean Minkowski spacetime, so we take the source and detector to be at rest on
average with respect to one another, at least initially, and so set $v^\mu = t^\mu$. The change in fractional
redshift between time $\tau_1$ and $\tau_2$, points $A$ and $B$ in Fig.~\ref{fig:spacetime},
may be written as
\begin{equation}
 \Delta\xi = \int_{\tau_1}^{\tau_2} d\tau \int_0^{\lambda_0}
 d\lambda \,
 R_{\alpha\beta\mu\nu}(\tau,\lambda)t^{\alpha}k^{\beta}t^{\mu}k^{\nu}\,.
\end{equation}
This expression gives $ \Delta\xi$ as an integral over the parallelogram in Fig.~\ref{fig:spacetime},
and can be interpreted as describing the change in a vector when parallel transported around a
closed path. 
Fluctuations of the Riemann tensor lead to fluctuations in this change in the redshift. Let
$\delta \xi$ be the variance in $\Delta \xi$, which may be expressed as a double integral of
the Riemann tensor correlation function, as
\begin{equation}
\delta \xi^2 = 
\int da \, da' \, C_{\alpha \beta \mu \nu \gamma \delta \sigma \rho} \;
t^\alpha k^\beta t^\mu k^\nu t^\gamma k^\delta t^\sigma k^\rho \,,
\label{eq:xi-int1}
\end{equation}
 where
 \begin{equation}
C_{\alpha \beta \mu \nu \gamma \delta \sigma \rho} = 
\langle R_{\alpha \beta \mu \nu}(x) R_{\gamma \delta \sigma \rho}(x') \rangle - 
\langle R_{\alpha \beta \mu \nu}(x)\rangle \langle R_{\gamma \delta \sigma \rho}(x') \rangle \,,
\end{equation}
and $da = d\tau d\lambda$.

We consider the geometry described in Sec.~\ref{sec:basic}, where the $t^{\mu}$ and $k^{\mu}$
are given by Eqs.~(\ref{eq:t}) and (\ref{eq:k}), respectively. Then we have 
\begin{equation}
R_{\alpha \beta \mu \nu}t^\alpha k^\beta t^\mu k^\nu = a^{-6}\;R_{uvuv} \,.
\end{equation}
Here the Riemann tensor undergoes passive fluctuations driven by stress tensor fluctuations,
so we set the Weyl tensor to zero and express  the Riemann tensor in terms of the Ricci tensor:
\begin{equation}
R_{uvuv} = \frac{1}{2} (g_{uu}\,R_{vv} + g_{vv}\,R_{uu} - 2g_{uv}\,R_{uv}) = \frac{1}{2} \, a^2 \, R_{uv} \,,
\end{equation}
where we use $g_{uu} = g_{vv} =0$ and $g_{uv} = -2 \, a^2$ for the spatially flat Robertson-Walker
metric in null coordinates. Now we may use the Einstein equation to
express the contracted Riemann tensor correlation function in terms of the stress tensor
 correlation function as
 \begin{equation}
I = C_{\alpha \beta \mu \nu \gamma \delta \sigma \rho} 
t^\alpha k^\beta t^\mu k^\nu t^\gamma k^\delta t^\sigma k^\rho =   
16 \pi^2 \ell_p^4\, a^{-4}(\eta_1) a^{-4}(\eta_2)\; C_{uvuv}(x_1,x_2) \,.
\end{equation}
Now we use Eqs.~(\ref{eq:weight}) and (\ref{eq:EM}) to write
 \begin{equation}
 C_{uvuv}(x_1,x_2) = a^{-2}(\eta_1)\, a^{-2}(\eta_2)\,  
C^{flat}_{uvuv}(x_1,x_2) =  \frac{4}{a^2(\eta_1)a^2(\eta_2)}(\partial_{u_1} \partial_{v_1} D)^2\,.
\end{equation}
As in the case of the luminosity fluctuations, we assume that the extent of the rays in the transverse 
spatial dimensions is small, and we may use Eq.~(\ref{eq:D}) to find
\begin{equation}
I = \frac{4 \ell_p^4}{\pi ^2 a^6(\eta_1) a^6(\eta_2)\; (u- i\epsilon)^4\, (v- i\epsilon)^4} \,.
\end{equation}

The  integrations in Eq.~(\ref{eq:xi-int1}) are over the area of the parallelogram in 
Fig.~\ref{fig:spacetime}, so $\int da = \int dt \int d\lambda$. We follow the procedure in 
Sec.~\ref{sec:lum} of changing variables to null coordinates,  approximating the integration over the
parallelogram by an integral over the rectangle of Fig.~\ref{fig:integ_area2}, and finally of
replacing the step functions which define the boundary of the rectangle by Lorentzian functions.
This leads to
 \begin{equation}
  \int dt_1 \int d\lambda_1 = 
 \tau\, L  \int_{-\infty}^{\infty} du_1\, f(u_1,\tau) \int_{-\infty}^{\infty} dv_1\,f(v_1,L) \; a^{3}(\eta_1) \,.
\end{equation}
Note that now we are integrating over the temporal direction, rather than averaging as before, which
leads to the overall factor of $\tau$. The expression for $\delta \xi^2$ may now be written as
\begin{equation}
\begin{split}
\delta \xi^2 = 
{} & \frac{4\ell^4_p}{\pi^2}\,\tau^2 \, L^2\, \int_{-\infty}^{\infty} du_1\, f(u_1,\tau) \int _{-\infty}^{\infty}du_2\, f(u_2,\tau) 
\int_{-\infty}^{\infty} dv_1f(v_1,L)  \int_{-\infty}^{\infty} dv_2 f(v_2,L)  \\
& \times a^{-3}(\eta_1) a^{-3}(\eta_2)\, \frac{1}{(u_1-u_2- i\epsilon)^4\,(v_1-v_2- i\epsilon)^4}  \,,
\end{split} \label{eq:-xi-int2}
\end{equation}
which is the analog of Eq.~(\ref{eq:-C-int3}), and contains the same powers of the scale factor.
This means that we may again use Eq.~(\ref{eq:expand}) and retain only the last term. The integrand
is now a function of $u=u_1-u_2$ and of $v=v_1-v_2$ only, so we may use Eq.~(\ref{eq:lorentzian_id})
to write
\begin{equation}
\delta \xi^2 = 
\frac{9 A^2 \ell_p^4}{\pi^4}\, \tau^2 \, L^2\,  \int_{-\infty}^{\infty} du \int_{-\infty}^{\infty} dv 
\frac{\mu}{u^2+\mu^2}\frac{\ell}{v^2 + \ell^2}  \frac{e^{\frac{i}{2}\omega (u+v)}}{(u- i\epsilon)^4\,(v- i\epsilon)^4} \,.
\end{equation}

The form is very close to that of the integral in Eq.~(\ref{eq:lum_int}), and is performed in the same
way, with the result
\begin{equation}
\delta \xi^2 = \frac{A^2 \ell_p^4}{2^{8} \pi^2 L^2 \tau^2}\,
\left[L \omega   \left(L^2 \omega ^2+6\right)+3e^{-L \omega }\right]
 \left[\tau \omega   \left(\tau^2 \omega ^2+6\right)+3e^{-\tau \omega }\right] \,.
\end{equation}
We are interested in the large $L$ limit, in which case $\delta \xi^2$ has the expected linear growth with
increasing travel distance $L$. If we also consider the limit that $\tau \gg 1/\omega$, then
we have that the root-mean-square  fractional redshift is given by
\begin{equation}
\delta \xi_{rms} \sim \sqrt{\delta \xi^2} \sim \frac{1}{16 \pi} A \,\ell_p^2\, \omega^3\; \sqrt{\tau\, L} \,.
\end{equation}
This result may be expressed
\begin{equation}
\delta \xi_{rms} \sim
4 \times 10^{-3} \, A\, \left(\frac{L}{\rm 1 Gpc}\right)^\frac{1}{2}  \, 
\left(\frac{\omega}{\rm 10 GeV}\right)^3\, \left(\frac{\tau}{\rm 1 hr}\right)^\frac{1}{2}\,.
 \label{eq:red-est}
\end{equation} 
Again, as in the case of luminosity fluctuations, the growth with increasing travel distance $L$ has 
produced a large effect at scales far from the Planck scale. The two cases differ in the power
dependence upon $L$, the extra power in Eq.~(\ref{eq:lum-est}) coming from the factor of $L^2$
in  Eq.~(\ref{eq:lum}). The two cases also differ in their dependence upon $\tau$. In the calculation
of  luminosity fluctuations, there is a temporal average over a time $\tau$, and the effect decreases
with increasing $\tau$. In the case of the redshift fluctuations, there is an integration over an interval
$\tau$, and the result grows as this time increases. This growth is analogous to the growth with 
increasing $L$. The result of integrating over the parallelogram in Fig.~\ref{fig:spacetime} increases
as the size of the parallelogram increases in either the spatial or the temporal direction. Just as
non-cancellation of anti-correlated fluctuations occurs in the $v$-direction, it also occurs in the
$u$-direction.  This growth in $\delta \xi^2$ with time may be related to a growth in relative motion of
the source and detector due to Riemann tensor fluctuations. This is a topic for future investigation.

\section{Variable Newton's Constant}
\label{sec:var-G}

In this section, we will recalculate both the variance in the fractional luminosity and the mean squared fractional redshift 
fluctuation, and obtain essentially the same results, but by means of a significantly different physical mechanism. We will 
return to Minkowski spacetime, so the scale factor is $a(\eta) = 1$, and instead postulate that the gravitational constant 
$G$ varies periodically with conformal time:
\begin{equation}
G(\eta) = G_0[1 + A\sin(\omega\eta)]\, ,
\label{eq:G}
\end{equation}
where $G_0 = \ell_p^2$ is understood to be the time-averaged value measured in the Cavendish experiment.
 Variation in $G$ is predicted by a wide class of scalar-tensor theories such as the Brans-Dicke theory~\cite{BD61}. 
For example, if we introduce a non-minimal scalar field $\varphi$ with a term in the action of the form
$\xi \, R \, \varphi^2$, where $R$ is the scalar curvature, then the effective Newton's constant becomes
dependent upon the value of $\varphi$. (See, for example, Ref. ~\cite{BV00}, Sect. 4.) If $\varphi$ is a classical field
with sinusoidal time dependence, then one can obtain a Newton's constant with a periodic time variation.
 
We present this avenue here as an illustration of the versatility of 
our core idea: that any introduction of a periodic time dependence into the calculation of observables calculated from the 
quantized EM field correlation function produces a linear growth of some kind.
The variable $G$ re-calculation of both the above luminosity and the redshift results is so similar that we need not repeat 
any 
of it here. We need only observe that, in this case, the Ricci tensor correlation function becomes
\begin{equation}
\begin{split}
C_{\alpha\beta\mu\nu} & = {} 64\pi^2\, G(\eta_1)G(\eta_2) \, \langle T_{\alpha\beta}T_{\mu\nu}\rangle \\
& = 64\pi^2\,G_0^2\,[1 + A\sin(\omega\eta_1)][1 + A\sin(\omega\eta_2)] \langle T_{\alpha\beta}\, T_{\mu\nu}\rangle \\
& = 64\pi^2\, G_0^2\,[1 + A\sin(\omega\eta_1) + A\sin(\omega\eta_2) + A^2\sin(\omega\eta_1)\sin(\omega\eta_2)]\,
\langle T_{\alpha\beta}T_{\mu\nu}\rangle \,.
\end{split}
\end{equation}
The final term, the product of $\sin$ functions, is identical to the product of $\sin$ functions that emerged from the Taylor 
expansions, at 2nd order in $A$, in the luminosity and redshift calculations, save for the factor of $9$. Without any further 
calculation, we can therefore conclude that the variable $G$ version of our results for  the variances $\delta \xi^2$ and
$\langle (\Delta \theta)^2\rangle$ are exactly one-ninth of the formulae previously calculated. Thus the 
root-mean-square fractional
luminosity  and redshift variations become $1/3$ of the results given in Eqs.~(\ref{eq:lum-est}) and  (\ref{eq:red-est}),
respectively. However in this case, the restriction that $A$ be small does not seem to be needed.

\section{Summary and Discussion}
\label{sec:sum}

In the previous sections, we have seen that time dependence, in either the scale factor or the gravitational
constant, can lead to non-cancellation of anti-correlated quantum fluctuations of the gravitational field. This
non-cancellation in turn leads to observable quantities which grow with increasing distance between a
source and a detector. The two observables which we considered are luminosity fluctuations and redshift
fluctuations, which are produced by fluctuations of the Ricci tensor and the Riemann tensor, respectively.
We consider a model in which both of these are passive fluctuations are driven by quantum fluctuations 
of a quantum stress tensor, specifically that for the quantized electromagnetic field in its vacuum state. 
In both cases, the dimensionless measure of fluctuations, the fractional luminosity or line width fluctuations,
is proportional to $\ell_p^2$, the square of the Planck length. Without the growth with increasing distance,
both quantities would be extremely small. However, the secular growth effect allows the possibility of
observable effects far from the Planck scale, as illustrated in Eqs.~\eqref{eq:lum-est} and \eqref{eq:red-est}.
In this paper, we have dealt only with integrals of stress tensor correlation functions, essentially second
moments of a probability distribution. Recent results on probability distributions for quantum stress tensor 
fluctuations~\cite{FFR10,FFR12}, indicate that the tails of these distributions fall off rather slowly, so large
fluctuations are not so rare as one might have expected. This opens the possibility of a different amplification
mechanism, which remains to be explored.

The time dependence which we have assumed is in the form of either an oscillating scale factor, 
Eq.~\eqref{eq:a-osc}, or an oscillating Newton's constant, Eq.~\eqref{eq:G}. The former can arise from an
$R^2$ term in the effective action for gravity, and the latter from an oscillating non-minimal scalar field. 
We have used a geometric optics approximation for both the luminosity and redshift fluctuation models,
so the frequency of the light rays used to probe the effects of the spacetime geometry fluctuations must be
somewhat higher than $\omega$, the oscillation frequency of the scale factor or of Newton's constant.  
We have integrated the relevant correlation functions over a two-dimensional spacetime region, with the extent
of the bundle of rays in the transverse spatial directions taken to be small. The integrations are over the regions 
depicted in Figs.~\ref{fig:integ_area1} and  \ref{fig:integ_area2}, with the latter an approximation to the former
for large travel distance. We have also used Lorentzian sampling functions to define to boundaries of the region
of integration for ease in computing integrals. 

If the distance between the source and detector is a cosmological distance, and the oscillation frequency is
sufficiently high, then  Eqs.~\eqref{eq:lum-est} and \eqref{eq:red-est} reveal that both luminosity and redshift 
fluctuations become potentially observable for light rays of even higher frequency. Now the term ``light ray''
means any relativistic particle. Typically, the luminosity effect is larger than the line broadening effect, due
to the extra power of $L$ in  Eq.~\eqref{eq:lum-est}. However, spectral line widths are easier to measure
accurately than are luminosity variations. In addition, the growth in  Eq.~\eqref{eq:red-est} with increasing
$\tau$ allows the possibility of enhancing  $\delta \xi_{rms}$ with longer observation times. It is also worth 
noting that  luminosity fluctuations are necessarily a passive effect, driven by stress tensor fluctuations, with
the resulting power of $\ell_p^2$ in Eq.~\eqref{eq:lum-est}. Line broadening, coming from a Riemann tensor 
correlation function, will also arise from  the active fluctuations of linearized quantum gravity. In this paper,
we have concentrated on the passive case, because the conformal transformation of the stress tensor greatly
simplifies the calculation. However, the analog of Eq.~\eqref{eq:red-est} for active fluctuations is expected
to be proportional to $\ell_p$, and hence describe a larger effect. This is a topic for further investigation.
In summary, we have shown that the mechanism of non-cancellation has the possibility to greatly amplify
quantum gravity effects, and potentially to lead to observable phenomena.

 \begin{acknowledgments}
 We would like to thank Ken Olum for useful discussions.
This work was supported in part by the National Science Foundation under Grant PHY-1205764.
\end{acknowledgments}

\end{document}